\documentclass[letter,oldversion]{aa}
\usepackage{graphicx}
\usepackage{txfonts}
\def\ltapprox{\raise 2pt \hbox {$<$} \kern-1.1em \lower 5pt \hbox {$\approx$}}

\def\ltsim{\; \raise0.3ex\hbox{$<$\kern-0.75em \raise-1.1ex\hbox{$\sim$}}\; }
\def\gtsim{\; \raise0.3ex\hbox{$>$\kern-0.75em \raise-1.1ex\hbox{$\sim$}}\; }

\def\ie{{\it i.e.,~}}

\def\eg{{\it e.g.,~}}
\def\egg{{\it e.g.~}}
\begin{document}
   \title{A morphological comparison between giant radio halos and radio mini--halos in galaxy clusters}

%   \subtitle{MH and GH in galaxy clusters}

   \author{R. Cassano\inst{1}\fnmsep\thanks{\email{rcassano@ira.inaf.it}},
	   M. Gitti\inst{2}, 
	   G. Brunetti\inst{1}}
\authorrunning{R. Cassano et al.}

   \offprints{R.Cassano}

   \institute{INAF - Istituto di Radioastronomia, via P. Gobetti 101,I-40129 Bologna, Italy\\
	\and
	INAF - Osservatorio Astronomico di Bologna,
via Ranzani 1, I-40127 Bologna, Italy}
%   \and
%   Dipartimento di Astronomia, Universit\`a di Bologna, via Ranzani 1, I-40127 Bologna, Italy

   \date{Received...; accepted...}

   \abstract{In this letter we present a morphological comparison
     between giant radio halos and radio mini-halos in galaxy
     clusters based on radio--X-ray luminosity, $P_{1.4}$-$L_{\rm X}$, and
     radio luminosity-size, $P_{1.4}$-$R_{\rm H}$, correlations.
     We report evidence that $P_{1.4}$-$L_{\rm X}$ and
     $P_{1.4}$-$R_{\rm H}$ trends may also exist for mini--halos: 
     mini--halo clusters share the same region of giant halo clusters in the $(P_{1.4},L_X)$ plane, whereas they are clearly separated in the $(P_{1.4},R_H)$ plane.
     The synchrotron emissivity of mini-halos is found to be more than $50$ 
     times larger than that of giant halos, implying a very efficient
     process for their origins. 
     By assuming a scenario of sporadical turbulent particle re-acceleration for both
     giant and mini halos, we discuss basic physical differences between these sources. Regardless of the origin of the turbulence, a more efficient source of injection of particles, which eventually takes part in the re-acceleration process, is required in mini-halos, and this may result from the central radio galaxy or from proton-proton collisions in the dense cool core regions.

}

\maketitle

\section{Introduction}

The intra--cluster medium (ICM) consists not only
of hot gas emitting in X-rays but also of non-thermal components.  
The major evidence for this comes from observations in the radio band
where Mpc--scale diffuse synchrotron emission from the ICM is detected in a number
of clusters (\egg Feretti 2005; Ferrari et al. 2008), 
indicating the presence of relativistic electrons and
magnetic fields.  These radio sources are generally referred to as
giant radio halos when located at the cluster center and radio relics
when located at the cluster periphery.  There are also some examples
of diffuse radio emission on smaller scales ($\sim 200$-$500$ kpc),
referred to as mini radio halos, extending around powerful radio 
galaxies at the center of some cool core clusters, \ie clusters characterized by a very peaked surface brightness profile and short central cooling time formerly known as
`cooling flow' clusters (\egg Peterson \& Fabian 2006).  
Galaxy clusters hosting giant halos are found to 
always be characterized by a peculiar dynamical status indicative of very
recent or ongoing merger events (\eg Buote 2001; Schuecker et al 2001),
whereas clusters hosting mini halos are characterized by a cool
core, with or without signs of moderate dynamical activity.   
Several statistical studies reveal that radio halos are not common in clusters 
(Giovannini et al. 1999; Kempner \& Sarazin 2001; Venturi et al. 2008; 
Brunetti et al. 2007; Cassano et al. 2008); instead, the statistics 
for mini halos is much poorer.

The main difficulty in understanding the origin of the synchrotron
emitting electrons in both giant halos and mini halos is
related to the fact that the diffusion length of the relativistic 
electrons is much shorter than the typical scale of the radio
emission (\eg Brunetti 2003). Therefore both giant halos and mini
halos cannot be explained in terms of simple 
diffusion of the
relativistic electrons from one or more cluster radio galaxies.  
Two main possibilities have been proposed so far to explain the origin
of both giant radio halos and mini radio halos:
{\it i)} {\it re-acceleration} models, whereby
relativistic electrons injected in the ICM are re-energized {\it in
  situ} by various mechanisms associated with turbulence in the ICM.
Turbulence in radio halos is supposed to be generated by massive merger events 
(\eg Brunetti et al. 2001; Petrosian et al. 2001). In mini radio halos,
 a seed large--scale turbulence frozen into the flow
could be amplified by the compression of the ICM in the cool core 
(Gitti et al. 2002, 2004);
{\it ii)} {\it secondary electron} models, whereby the
relativistic electrons are secondary products of the hadronic
interactions of cosmic rays (CR) with the ICM (\eg Dennison 1980;
Blasi \& Colafrancesco 1999, Pfrommer \& En\ss lin 2004).
Although the properties of giant halos and mini
halos are clearly different (different size, different
dynamical state of the hosting clusters), it is not
clear whether they are different astrophysical
phenomena or if they might share similar physics. 
In this letter we carry out a morphological comparison between giant 
radio halos and mini radio halos aimed at studying the differences between their
physical properties.  
We also consider the case of diffuse cluster
sources with intermediate properties between giant halos and mini halos.
To do this we investigate the presence of
scaling relations between the main properties of these sources.
A $\Lambda$CDM cosmology ($H_{o}=70\,\rm km\,\rm s^{-1}\,\rm Mpc^{-1}$, $\Omega_{m}=0.3$,
$\Omega_{\Lambda}=0.7$) is adopted.

\section{Selection of clusters with diffuse radio emission}

We collect from the literature all clusters with well-studied giant radio halos
and mini radio halos, as defined in the following.

\noindent $\bullet$ {\it Giant radio halos} (hereafter {\bf GHs}):
radio sources typically extended on Mpc scales found at the center of
non-cool core clusters. We consider as belonging to this class all the 
diffuse radio sources larger than the core radius of the
cluster. This is an important physical information relating the origin
of these sources to physical processes taking place on cluster scales.
We end up with 19 GHs, 15 of which have already been analyzed in Cassano et
al. (2007, hereafter C07), 3 come from a recent survey carried out with the GMRT: A\,209, A\,697 and RXCJ\,2003.5-2323 (Venturi et al. 2008), and 1 is a smaller halo
found in A\,3562 (Venturi et al. 2003; Giacintucci et al. 2005), which extends
however beyond the cluster core. 
The presence of recent or ongoing merger events is well-established 
in all clusters of galaxies hosting these GHs.

\noindent $\bullet$ {\it Mini radio halos} (hereafter {\bf MHs}):
radio sources with total size $\sim 140-500\,h_{70}^{-1}$ kpc
surrounding the central radio galaxy at the center of cool core
clusters (hereafter {\bf CCCs}; Tab.~\ref{Tab.minihalo}). 
Their size is comparable to that of the cooling region and smaller than the
cluster core. For a long time, clusters hosting MHs have been
identified with relaxed clusters based on the presence of cool cores
at their center; however, signs of minor mergers and/or accretion of
small subclumps have recently been detected in 5 clusters with MHs
thanks to Chandra and XMM--Newton observations
(Perseus: Churazov et al. 2004; A\,2626: Wong et al. 2008;
RX J1347.5$-$1145: Allen et al. 2002; Gitti \& Schindler 2004, Gitti et
al. 2007a, 2007b; A\,2390: Allen et al. 2001; Z\,7160: Mazzotta \& Giacintucci 2008).

\noindent 
For completeness we also consider the case of smaller scale radio sources 
in some merging clusters: A\,2218 and A\,2142 (Giovannini \& Feretti 2000), and  RXCJ1314.4-2515 (Venturi et al. 2007), which do not belong to the two classes above.

\section{Observed scaling relations}

There are several observed correlations for GHs that relate
thermal and non-thermal properties of the ICM: those between the radio
power at 1.4 GHz, $P_{1.4}$, and the X-ray luminosity, $L_{\rm X}$,
temperature, $T$, and cluster mass, $M$ (\eg Liang et al. 2000;
Feretti 2000; Govoni et al. 2001; Cassano et al. 2006).
Additional correlations were also explored between $P_{1.4}$ 
and the size of GHs, $R_{\rm H}$, the total cluster mass within $R_{\rm H}$,
and the cluster velocity dispersion (C07). 
C07 find a trend between $R_{\rm H}$ and the cluster
virial radius, $R_{\rm v}$, and show that all the other correlations 
explored so far can be derived by combining the $R_{\rm
  H}$-$R_{\rm v}$ and $P_{1.4}$-$R_{\rm H}$ scalings.
This suggests that there are essentially two main scaling
relations that carry the leading information on the physics of
non-thermal components.
The statistical properties of MHs are less explored, indeed few
objects are known to possess a MH and only a trend between $P_{1.4}$
and the maximum power of the cooling flow (estimated as $P_{\rm
  CF}=\dot{M}\,kT/\mu m_p$, Gitti et al. 2004) has been found so far
for this class of sources.

In Fig.\ref{Fig.Lr_Lx} we show the distribution of GHs in the 
($P_{1.4},L_{\rm X}$) \footnote{ 
Radio powers that have been measured at different frequencies are converted at 1.4 GHz 
with a spectral index $\alpha=1.3$, with $S\propto \nu^{-\alpha}$.}
plane together with the correlation from Cassano et al. (2006). 
The MHs are also reported in Fig.\ref{Fig.Lr_Lx}, where we find that they share the same region of GHs and that a possible $P_{1.4}$-$L_{\rm X}$ correlation also exists for MHs and is barely consistent with that of GHs. 
In Fig.\ref{Fig.Lr_Lx} we also report upper limits to the radio power of 
CCCs without detected MHs taken from the statistical sample of X-ray luminous clusters with deep radio follow up of Venturi et al. (2008).  
These CCCs without MHs are all relaxed clusters with a central active radio galaxy.
Limits were obtained following the approach given in Brunetti et al. (2007, hereafter B07) and lie one order of magnitude below the radio power of MHs. Despite the poor statistics, this suggest that MHs may not be common in CCCs.

In Fig.~\ref{Fig.Lr_RH} we show the distribution of GHs in the 
($P_{1.4},R_{\rm H}$) plane, where $R_{\rm H}$ is the radius of
the emitting region (as in C07). Following the procedure in C07,
we also derived $R_H$ for MHs from the radio images of these sources.
We also find a trend between $P_{1.4}$ and $R_H$ for MH ( 
Fig.~\ref{Fig.Lr_RH}) with $P_{1.4}$ rapidly increasing in larger MHs.
This trend, however, is not consistent with what is found for GH. 
The clear separation demonstrates the importance of exploring the 
distributions of these radio sources in different planes
to distinguish between different radio sources
and to investigate their physics and origins.

\noindent In Figs. 1 \& 2
we also report on A\,2218, A\,2142, and RXCJ1314.
They are less luminous than GH (Bacchi et al. 2003) and MH
in clusters with similar $L_X$, and are located between GH and MH
in the ($P_{1.4},R_{\rm H}$) plane. 
They are in merging clusters and are smaller (smaller emitting volumes and
thus radio powers) than GH, suggesting that they might be 
GH at an early evolutionary stage.

\label{scalings}
\begin{table}
\caption{Sample of galaxy clusters hosting mini radio halos.}
\begin{center}
\label{Tab.minihalo}
\begin{tabular}{lclll}
\hline\noalign{\smallskip}
Name   & z &  Log $L_{\rm X}$ & Log $P_{\rm 1.4\,GHz}$ & Log $R_{\rm H}$\\
       &   &  (erg/s)& (W/Hz) & (kpc)\\
\noalign{\smallskip}
\hline\noalign{\smallskip}
Perseus (a) & 0.018 & $44.82^{+0.01}_{-0.01}$ & $24.27^{+0.04}_{-0.04}$ & 2.12 \\
A2390  (b) & 0.228 & $45.13^{+0.05}_{-0.05}$ & $24.99^{+0.02}_{-0.02}$ & 2.26 \\
A2626  (c) & 0.060 & $44.03^{+0.05}_{-0.05}$ & $23.36^{+0.03}_{-0.10}$ & 1.85 \\
RX J1347.5$-$1145 (d) & 0.451 & $45.65^{+0.05}_{-0.05}$ & $25.28^{+0.01}_{-0.01}$ & 2.41 \\
Z7160 (e) & 0.258 & $44.93^{+0.03}_{-0.03}$ & $24.34^{+0.05}_{-0.05}$ & 2.24 \\
RBS 797 (f)   & 0.350  & $45.31^{+0.02}_{-0.02}$ & $24.63^{+0.04}_{-0.04}$  & 2.01 \\
\noalign{\smallskip}
\hline
\end{tabular}
\end{center}
References for the radio data in brackets: 
(a) Sijbring 1993,
(b) Bacchi et al. 2003,
(c) Gitti et al. 2004, 
(d) Gitti et al. 2007, 
(e) Venturi et al. 2008,
(f) Gitti, Feretti \& Schindler 2006, Gitti et al., in prep.
\end{table}

\section{Basic interpretation of the observed scalings}

\begin{figure}
\begin{center}
\includegraphics[width=0.33\textwidth]{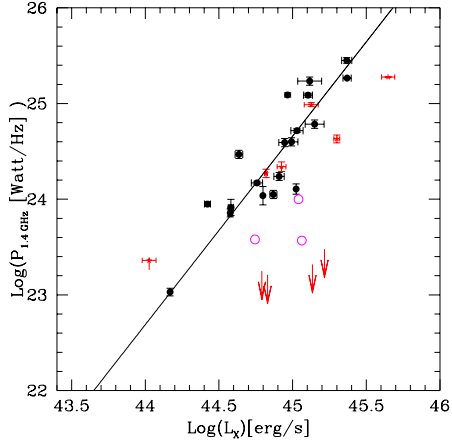}
\caption[]{Radio power at 1.4 GHz versus X-ray luminosity 
in the [0.1-2.4] keV band of clusters with GHs (black circles), MHs (red asterisks),
and small-scale radio emissions (magenta open circles). 
Arrows are upper limits to the radio power
of CCCs without MHs (see text).
The black solid line is the best-fit correlation for GHs 
(from Cassano et al. 2006). }
\label{Fig.Lr_Lx}
\end{center}
\end{figure}

In the case of GHs B07 found that 70\% of clusters in the Venturi et al. (2008) sample are radio quiet, not showing Mpc scale synchrotron radio emission, and that the limits on their radio powers lie one order of magnitude below the correlation followed by clusters with GHs. 
They discussed that the bi-modality between radio quiet clusters and GH clusters is in line with the re-acceleration scenario, in which turbulence powers up GHs only for a limited period
during cluster mergers, and disfavors secondary models that would instead predict GHs that 
are much more common in galaxy clusters. 
Despite the poor statistics, MHs in Fig.\ref{Fig.Lr_Lx} show a behavior similar to 
that of GHs in B07. Furthermore, these MHs are characterized by non-relaxed cores, whereas
the CCCs without MHs are all relaxed clusters. 
This might suggest that turbulence connected with merger activity
could also play a role in the acceleration of electrons in MHs. 
More generally,
turbulence necessary to trigger MHs could result
from the amplification in the cool core of a seed turbulence 
present in the ICM (Gitti et al. 2002, 2004), could be connected with the 
gas-sloshing mechanism in CCCs (Mazzotta \& Giacintucci 2007), 
or could be driven by minor mergers (see also Gitti et al. 2007a) that are common in
CCCs with MHs (Sect.~2).
Another possibility is that turbulence takes a small fraction 
of the energy released by the ``bubbles'' rising from the central AGN, and this can offset the cooling in most clusters (\eg McNamara \& Nulsen 2007). 
All 6 MH clusters indeed have an active radio galaxy at
their center and 5 have cavities in the X-ray ICM.

\begin{figure}
\begin{center}
\includegraphics[width=0.33\textwidth]{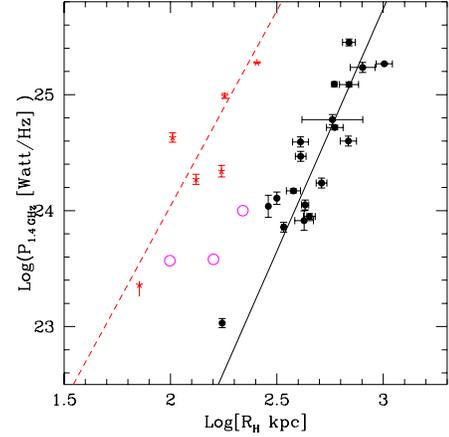}
\caption[]{Radio power at 1.4 GHz versus radio size of GHs (black circles) 
and MHs (red asterisks), and small-scale radio emissions (magenta open circles). 
The black solid line and the red dashed line are 
the best-fit correlations for GHs ($P_{1.4}\propto R_{\rm H}^{4.18}$, 
from Cassano et al. 2007) and for MHs 
($P_{1.4}\propto R_{\rm H}^{3.4}$), respectively.}
\label{Fig.Lr_RH}
\end{center}
\end{figure}

A detailed physical modeling of MHs in CCCs is beyond
the scope of this letter, but we can derive some basic
constraints here on their physical parameters.
We find that, although emitting a similar radio power,
the radius of MHs is typically a factor $\approx 4$ smaller than
that of GHs (see Fig.~\ref{Fig.Lr_RH}). This implies a synchrotron emissivity 
for MHs $\approx 50$ times larger than that of GHs\footnote{This value should be considered as a lower limit since MHs have also more peaked profiles than GHs.}. 
Regardless of the origin of the emitting electrons, the ratio
between the synchrotron emissivity of MHs ($\dot{\varepsilon}_{\rm MH}$)
and of GH ($\dot{\varepsilon}_{\rm GH}$) can be written as

\begin{equation}
\frac{\dot{\varepsilon}_{\rm MH}}{\dot{\varepsilon}_{\rm GH}}= 
\frac{n_{\rm rel}^{\rm MH}}{n_{\rm rel}^{\rm GH}}
\bigg(\frac{B_{\rm MH}}{B_{\rm GH}}\bigg)^{\alpha+1}
\label{Eq.1}
\end{equation}

\noindent where $n_{\rm rel}^{\rm MH}$ and $B_{\rm MH}$ ($n_{\rm
  rel}^{\rm GH}$ and $B_{\rm GH}$) are the number density of radio
emitting electrons (at the energy needed to emit the
observed synchrotron radiation) and the mean value of the magnetic field strength
within the MH (GH), respectively, and $\alpha$ is the radio spectral
index of the synchrotron spectrum, which is similar in
GHs and MHs ($\alpha\sim 1.1-1.3$, 
\eg Feretti \& Giovannini 2007).  
The measure of $B$ in the ICM is quite problematic and 
different methods often give different estimates.
Faraday rotation measure studies generally found a few to 10 $\mu$G in non-CCCs and $\sim10-30\,\mu$G in the central region of CCCs 
(Clarke 2004, Govoni \& Feretti 2004), whereas methods based on inverse Compton emission found from $\sim 0.1$ to $\mu$G (Fusco-Femiano et al. 2004; Sanders et al. 2005).  
There is, however, agreement on the fact that the magnetic field at the center of 
CCCs is larger than that on the Mpc scale in non-CCCs. 

A suitable assumption for the ratio $(B_{\rm MH}/B_{\rm GH})$ in Eq.\ref{Eq.1} (with
$B_{\rm MH}>>B_{\rm GH}$) allows us to reproduce the observed ratio between the emissivities.
However, the difference in terms of $B$ cannot be the only cause of the large
synchrotron emissivity in MHs. Indeed a large $B$ in CCCs produces a fast cooling of relativistic electrons due to synchrotron losses (that make $n_{\rm rel}^{\rm MH}$ smaller), and this implies the important point that a very efficient mechanism of injection and/or
acceleration of relativistic electrons should also be active in MHs.

\begin{figure}
\begin{center}
\includegraphics[width=0.3\textwidth]{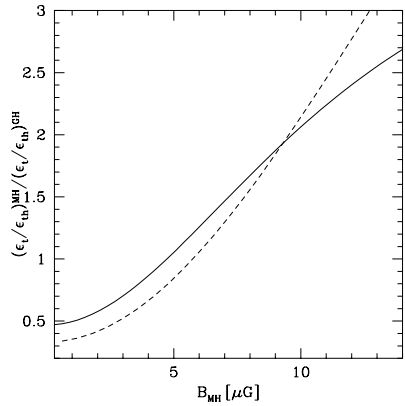}
\caption[]{Ratio between turbulence energy densities of MHs and GHs normalized 
to the thermal ones as a function of $B_{MH}$. The calculations are reported 
for $f\approx1$, $z=0.1$, $T_{GH}/T_{MH}\approx 3$ and 
%$R_H^{GH}/R_H^{MH}\approx 3.5$,
in the case of $B_{MH}\approx 3 B_{GH}$ (solid line) and in the case
$B_{MH}\approx 6 B_{GH}$ (dashed line).}
\label{Fig.stupido}
\end{center}
\end{figure}

To quantify this point in a relevant case, we assume that electrons are re-accelerated 
sporadically by turbulence injected in the emitting region by some process.
Electrons are accelerated up to the energy where acceleration
is balanced by losses, $m_{\rm e} c^2 \gamma_{\rm b} \propto \chi/\beta$, where $\chi$ is the acceleration efficiency and $\beta=(B_{\rm cmb}^2 + B^2)$ accounts for the synchrotron and inverse Compton losses\footnote{$B_{\rm cmb}\approx 3.2(1+z)^2\,\mu$G is the equivalent magnetic field strength of the cosmic microwave background (CMB) radiation.}, 
and a corresponding break forms in the emitted synchrotron
spectrum at $\nu_b \propto B \gamma_b^2$.
Magnetosonic waves are proposed as possible sources of particle acceleration in the
ICM (Cassano \& Brunetti 2005; Brunetti \& Lazarian 2007);
and in this case, following C07, the synchrotron emissivity 
(if the damping of turbulence is dominated by thermal electrons, for $\epsilon_{\rm rel}/\epsilon_{\rm th} << 1$) is

\begin{equation}
\dot{\varepsilon}_{\rm syn} \propto \dot{\varepsilon}_{\rm t} \, 
(\epsilon_{\rm rel}/{\epsilon_{\rm th})}\, B^2\, T^{1/2}\,\beta^{-1} 
\label{Eq.2}
\end{equation}

\noindent where $\dot{\varepsilon}_{\rm t}$ is the turbulence
injection rate, $\epsilon_{\rm rel}/\epsilon_{\rm th}$ the ratio
between the energy density of relativistic and thermal particles, and
$T$ the cluster temperature. 
We consider the case in which GH and MH have $\nu_{\rm b}^{\rm MH}= f\,\nu_{\rm b}^{\rm GH}$, 
and $f=1$ would imply that GH and MH have similar spectral index, in line with
present observations. 
From Eqs.~35 and 36 in Cassano \& Brunetti (2005) (in the case 
$\epsilon_{\rm rel}/\epsilon_{\rm th} << 1$) $\dot{\varepsilon_{\rm t}}\propto 
\beta\,n_{\rm th}(T\,\nu_{\rm b}/B)^{1/2}$ ($n_{\rm th}$ is the thermal gas density)
and assuming $\nu_{\rm b}^{\rm MH}= f\,\nu_{\rm b}^{\rm GH}$, one can derive the ratio
between the turbulence injection rate in the two populations of sources.
A comparison between the energy injected in the form of magnetosonic waves (in unit of the thermal energy, $\epsilon_t/\epsilon_{th}$) in GH and MH (assuming $\epsilon_t=\dot{\varepsilon}_{\rm t}\Delta\tau\approx\dot{\varepsilon}_{\rm t} R_H/c_s$,
with $\Delta\tau$ and $c_s$ being the injection time scale and the sound speed, respectively)
is reported in Fig.~\ref{Fig.stupido}. 
This figure shows that the energetic request in the case of MH is comparable to that of 
GH provided that the magnetic field in CCCs is not significantly larger than
$\approx 10\,\mu$G; in this case, $\epsilon_t/\epsilon_{th}
\sim 0.1-0.3$ is typically required to have $\nu_b \approx 1$ GHz 
(Brunetti \& Lazarian 2007).

\noindent  
By combining the expression for $\dot{\varepsilon}_{\rm t}^{\rm MH}/\dot{\varepsilon}_{\rm t}^{\rm GH}$ with Eq.2 one finds the ratio between the synchrotron emissivities in GH and MH:

\begin{equation}
\frac{\dot{\varepsilon}_{\rm syn}^{\rm MH}}{\dot{\varepsilon}_{\rm syn}^{\rm GH}}=
f^{1/2}\,\big(\frac{B_{\rm MH}}{B_{\rm GH}}\big)^{3/2}\times
\frac{\epsilon_{\rm rel}^{\rm MH}}{\epsilon_{\rm rel}^{\rm GH}}\,\rm.
\label{Eq.4}
\end{equation}

\noindent 
This implies that to understand the large differences in terms of emissivity 
($\dot{\varepsilon}_{\rm syn}^{\rm MH}>50\,\dot{\varepsilon}_{\rm syn}^{\rm GH}$)
observed between MH and GH an extra amount of relativistic (fossil) electrons
that take part in the acceleration process in CCCs is necessary (
$\epsilon_{\rm rel}^{\rm MH}/\epsilon_{\rm rel}^{\rm GH}\approx
4$-$10$ assuming $B_{\rm MH}/B_{\rm GH}\approx 3-6$ and $f\approx1$).
This extra amount of relativistic electrons may be provided naturally by the central radio galaxy or by the p-p collisions (CR protons
and thermal protons) in the dense CCCs that may inject efficiently secondary electrons,
to be re-accelerated during the activity of MH.

\section{Conclusions}

In this letter we have compared the observed properties of mini radio halos
(MHs) and giant radio halos (GHs) in clusters of galaxies.  GHs are
the most prominent evidence of non-thermal components in the ICM
and several correlations between thermal and non-thermal properties have been
explored for these sources, including those relating 
$P_{1.4}$ to $L_{\rm X}$ and to $R_{\rm H}$ (\eg Cassano et
al. 2006, 2007; Brunetti et al. 2007).
On the other hand, 
an extensive investigation of the statistical properties of MHs is
presently not possible since only a few clusters host well-studied
MHs. 

We collected a sample of MH and compared their behavior 
with that of GHs in the ($P_{1.4},L_{\rm X}$) and 
($P_{1.4},R_{\rm H}$) planes. 
We find that $P_{1.4}-L_{\rm X}$ and $P_{1.4}-R_{\rm H}$ 
trends may also exists for MHs.
While in the ($P_{1.4},L_{\rm X}$) plane MHs and GHs share the same region,
in the ($P_{1.4},R_{\rm H}$) plane MHs do not follow the same correlation of GHs 
at smaller radii, but are clearly separated.
We find that the typical synchrotron emissivity of
MHs is at least $50$ times larger than that of GHs.
This implies a very efficient mechanism at the origin of the emitting electrons
in MHs.  
For completeness we also consider the few cases of smaller scale emission in
non-CC (and without central radio galaxy) merging clusters. These sources are
morphologically intermediate between GH and MH and may be GH at some early evolutionary
stage.

The distribution in the ($P_{1.4},L_{\rm X}$) plane of a
small sample of CCCs with available radio
observations suggests that MHs are not ubiquitous in CCCs,
with upper limits for CCCs without diffuse radio emission
well below the radio power of MHs in clusters with similar $L_X$.
Those CCCs without MHs also appear to be more relaxed than that
with MHs. All these findings, if confirmed, would point in favor
of sporadic turbulent re-acceleration as the origin of the emitting particles.
In addition to the possibilities already
explored in the literature (Gitti et al. 2002; Mazzotta \& Giacintucci
2007), minor mergers (see also Gitti et al. 2007a) and/or the central AGN outbursts 
may contribute to the injection of turbulence in the ICM of CCCs.
By adopting this scenario, under the assumption that magnetosonic waves drive the particle acceleration process, we find that 
the larger synchrotron emissivity of MHs can be explained by assuming that the energy density of the relativistic particles that interact with turbulence is 
about one order of magnitude higher than in GHs, and that this does not necessarily
imply a larger amount of turbulence in MHs.
The extra amount of relativistic particles in these sources may be provided by the central cluster galaxy or by secondary electrons injected in the dense cool core region.

\begin{acknowledgements}
  We thank G. Setti, S. Giacintucci, T. Venturi, and the anonymous referee for useful comments.
  This work is partially supported by grants PRIN-INAF2005, and ASI-INAF I/088/06/0.
\end{acknowledgements}


\begin{thebibliography}{}

\bibitem{} Allen, S. W.; Ettori, S.; Fabian, A. C., 2001, MNRAS 324, 877
\bibitem{} Allen, S. W.; Schmidt, R. W.; Fabian, A. C., 2002, MNRAS 335, 256
\bibitem{} Bacchi M., Feretti L., Giovannini G., Govoni F., 2003, A\&A, 400, 465
\bibitem{} Blasi P., Colafrancesco S., 1999, APh 12, 169
\bibitem{} Brunetti G., 2003, in ``Matter and Energy in Clusters of Galaxies'', ASP Conf. Series, vol.301, p.349,eds. S. Bowyer and C.-Y. Hwang.
\bibitem{} Brunetti G., Setti G., Feretti L., Giovannini G., 2001, MNRAS 320, 365
\bibitem{} Brunetti G., Venturi T., Dallacasa D. et al. 2007, ApJ Letter 670, 5
\bibitem{} Brunetti G., Lazarian A., MNRAS 378, 245
\bibitem{} Buote D.A, 2001, ApJ 553, 15
\bibitem{} Cassano R., Brunetti G. 2005, MNRAS 357, 1313
\bibitem{} Cassano R., Brunetti G., Setti G., 2006, MNRAS 369,1577
\bibitem{} Cassano R., Brunetti G., Setti G. et al. 2007, MNRAS 378, 1565
\bibitem{} Cassano, R.; Brunetti, G.; Venturi, T. et al. 2008, A\&A 480, 687
\bibitem{} Churazov, E.; Forman, W.; Jones, C. et al. 2004, MNRAS 347, 29
\bibitem{} Clarke T.E. 2004, JKAS 37, 337
\bibitem{} Dennison B., 1980, ApJ 239L
\bibitem{} Feretti L., 2000, Invited review at IAU 199 `The Universe at Low Radio Frequencies' in Pune, India, 1999
\bibitem{} Feretti L., 2005, in 'X-Ray and Radio Connections', published electronically by NRAO, eds. L.O.Sjouwerman and K.K.Dyer
\bibitem{} Feretti, L., Giovannini, G., 2007, published in Springer Lecture Notes in Physics, ``Panchromatic View of Clusters of Galaxies and the Large-Scale Structure'', Editors M. Plionis, O. Lopez-Cruz, and D. Hughes
\bibitem{} Ferrari F., Govoni F., Schindler S. et al. 2008, SSRv 134, 93
%Space Science Reviews, astro-ph/0801.0985
\bibitem{} Fusco-Femiano R., Orlandini M., Brunetti G., et al. 2004, ApJ 602, 73
\bibitem{} Giacintucci, S.; Venturi, T.; Brunetti, G. et al. 2005, A\&A 440, 867
\bibitem{} Giovannini G., Tordi M., Feretti L., 1999, NewA 4, 141
\bibitem{} Gitti M., Brunetti G., Setti G., 2002, A\&A 386, 456
\bibitem{} Gitti, M.; Brunetti, G.; Feretti, L.; Setti, G., 2004,
A\&A 417, 1
\bibitem{} Gitti M., Schindler S., 2004, A\&A 427, L9
\bibitem{} Gitti M., Feretti L., Schindler S., 2006, A\&A 448, 853
\bibitem{} Gitti M., Ferrari C., Domainko W. et al. 2007a, A\&A 470L, 25
\bibitem{} Gitti M., Piffaretti, R.; Schindler, S., 2007b, A\&A 472, 383
\bibitem{} Govoni F., Feretti L., Giovannini G. et al. 2001, A\&A, 376, 803
\bibitem{} Govoni F., Feretti L., 2004, Int. J. Mod. Phys. D 13, 1549
\bibitem{} Kempner J.C., Sarazin C.L., 2001, ApJ 548, 639
\bibitem{} Liang H., Hunstead R.W., Birkinshaw M., Andreani P., 2000, ApJ 544, 686
\bibitem{} Mazzotta, P.; Giacintucci, S., 2008, ApJ Letter 675, 9
\bibitem{} McNamara B.R., Nulsen P.E.J., 2007, ARA\&A 45, 117
\bibitem{} Peterson J. \& Fabian A.C., 2006, Phys. Rep., 427, 1
\bibitem{} Petrosian V., 2001, ApJ 557, 560 
\bibitem{} Pfrommer C., En\ss lin T. A. 2004, A\&A 413, 17
\bibitem{} Sanders, J.S.; Fabian, A.C.; Dunn, R.J. H. 2005, MNRAS 360, 133
\bibitem{} Schuecker P., B\"ohringer H.; Reiprich T.H., et al. A\&A 378, 408
\bibitem{} Sijbring D. 1993, A radio Continuum and HI Line Study of the Perseus Cluster, Ph.D. Thesis, Groningen
\bibitem{} Venturi, T.; Bardelli, S.; Dallacasa, D. et al. 2003, A\&A 402, 913
\bibitem{} Venturi T., Giacintucci S., Brunetti G. et al. 2007, A\&A 463, 937
\bibitem{} Venturi T., Giacintucci S., Dallacasa D. et al. 2008, A\&A 484, 327	
\bibitem{} Wong K.-W., Sarazin, C.L., Blanton E.L., Reiprich, T. H. 2007, AJ in press, 
astro-ph/0803.1680

\end{thebibliography}
\end{document}